\documentclass[10pt, conference] {IEEEtran}
\usepackage[pdftex]{graphicx}
\graphicspath{{./graphics/}{./matlab_plots/}}
\usepackage{epstopdf}
\usepackage{amsmath}
\usepackage{amssymb}
\usepackage{acronym}
\usepackage{xcolor}

\newcommand{\bones}{\mathbf{1}}
\newcommand{\bzeros}{\mathbf{0}}

\newcommand{\bsigma}{\mbox{\boldmath{$\sigma$}}}

  \newcommand{\bs}{\mathbf{s}}

  \newcommand{\bw}{\mathbf{w}}

  \newcommand{\bx}{\mathbf{x}}

  \newcommand{\by}{\mathbf{y}}

  \newcommand{\bF}{\mathbf{F}}

  \newcommand{\bH}{\mathbf{H}}

  \newcommand{\bI}{\mathbf{I}}

  \newcommand{\bP}{\mathbf{P}}

  \newcommand{\bU}{\mathbf{U}}

  \newcommand{\bV}{\mathbf{V}}

  \newcommand{\bSigma}{\mathbf{\Sigma}}

  \newcommand{\und}{\underline}

  \newcommand{\bFrf}{\mathbf{F}_{\mathrm{RF}}}
  \newcommand{\bFbb}{\mathbf{F}_{\mathrm{BB}}}

\newcommand{\bFc}{\mathbf{F}_{\mathrm{C}}}
  \newcommand{\bFopt}{\mathbf{F}_{\mathrm{opt}}}

  \newcommand{\ubP}{\underline{\bP}}

\newtheorem{lemma}{Lemma}

\newtheorem{theorem}{Theorem}

\newcommand{\beq}{\begin{equation}}

\newcommand{\enq}{\end{equation}}

\newcommand{\beqa}{\begin{eqnarray}}

\newcommand{\enqa}{\end{eqnarray}}

\newcommand{\bea}{\begin{array}}

\newcommand{\ena}{\end{array}}

\newcommand{\bef}{\begin{figure}}

\newcommand{\enf}{\end{figure}}

\newcommand{\bds}{\begin {itemize}}

\newcommand{\eds}{\end {itemize}}

\newcommand{\bdf}{\begin{definition}}

\newcommand{\blm}{\begin{lemma}}

\newcommand{\edf}{\end{definition}}

\newcommand{\elm}{\end{lemma}}

\newcommand{\bthm}{\begin{theorem}}

\newcommand{\ethm}{\end{theorem}}

\newcommand{\cC}{{\ensuremath{\mathcal{C}}}}

\newcommand{\cI}{{\ensuremath{\mathcal{I}}}}

\newcommand{\cN}{{\ensuremath{\mathcal{N}}}}

\renewcommand{\vec}{\mbox{vec}}

\newcommand{\sigmax}[1]{\sigma_{\mathrm{#1}}}
\newcommand{\Nx}[1]{N_{\mathrm{#1}}}
\newcommand{\bFx}[1]{\mathbf{F}_{\mathrm{#1}}}
\newcommand{\thetax}[2]{\theta_{\mathrm{#1}}^{\mathrm{#2}}}
\newcommand{\phix}[2]{\phi_{\mathrm{#1}}^{\mathrm{#2}}}

\acrodef{MIMO} {Multiple Input Multiple Output}
\acrodef{RFPN} {RF Phase-shifting Network}
\acrodef{BS} {base station}
\acrodef{UE} {user equipment}
\acrodef{M-MIMO} {Massive MIMO}
\acrodef{HBF} {Hybrid Beamforming}
\acrodef{RFBN} {RF Beamforming Network}
\acrodef{SNR} {Signal to Noise Ratio}
\acrodef{SINR} {Signal to Interference Plus Noise Ratio}
\acrodef{LTE} {Long Term Evolution}
\acrodef{FFT} {Fast {F}ourier Transform}
\acrodef{OFDM} {Orthogonal Frequency Division Multiplexing}
\acrodef{FRM} {Frequency Response Masking}
\acrodef{CPG} {Cancellation Pulse Generator}
\acrodef{FIR} {Finite-length Impulse Response}
\acrodef{LS} {Least Squares}

\makeatother
\begin{document}

\title{Hybrid Analog and Digital Precoding: From Practical RF System Models to Information Theoretic Bounds}
\author{
\IEEEauthorblockN{ Vijay Venkateswaran, Rajet Krishnan}
\IEEEauthorblockA{Huawei Technologies, Sweden, \{vijay.venkateswaran, \, rajet.krishnan\}@huawei.com}
}

\maketitle

\begin{abstract}
Hybrid analog-digital precoding is a key millimeter wave access technology, where an antenna array with reduced number of radio frequency (RF) chains is used with an RF precoding matrix to increase antenna  gain at a reasonable cost. However, digital and RF precoder algorithms  must be accompanied by a detailed system model of the RF precoder. 
In this work, we provide fundamental RF system models for these precoders, and show their impact on achievable rates. We show that hybrid precoding systems suffer from significant degradation, once the limitations of RF precoding network are accounted. We subsequently quantify this performance degradation, and use it as a reference for comparing the performance of  different precoding methods. These results indicate that hybrid precoders must be redesigned (and their rates recomputed) to account for practical factors.

\end{abstract}
\section{Introduction}

Massive-\ac{MIMO} and millimeter wave access are two of the most promising directions for next generation wireless systems. Both techniques employ large scale antenna arrays to either spatially separate users or to increase the gain of the transmitted signals \cite{Marzetta:large_scale_mimo, Roh:mmwave}. Significant increase in the number of transceivers has rendered the cost of large-scale antenna arrays to be prohibitively expensive. In order to bring down this cost, \emph{hybrid precoders} have been proposed \cite{Elayach:hybrid, Alkhateeb:hybrid, Vijay:hybridmtt, Molisch:RF_phase_shifter}. In these systems, a reduced number of transceivers and digital baseband units are connected to the large antenna array through a \ac{RFPN}. In such a setup, part of the precoding is performed in digital baseband followed by rest of beamforming in RF using a bank of phase shifters. Reducing the number of digital baseband units and transceiver chains brings down the cost \cite{Wright:lsas}, while the \ac{RFPN} ensures that base station is able to focus signals towards specific users. Modeling and implementation of digital baseband network is trivial, whereas RF system modeling and implementation is not straightforward, and their impact has not been explored.

\subsection{RF System Perspective}

Most of the existing literature on hybrid precoders \cite{Elayach:hybrid, Alkhateeb:hybrid,Lau:hybrid}  oversimplify the practical constraints involved in the design of RF precoder, focus mostly on digital algorithms and disregard the practical challenges that one would typically encounter while designing an \ac{RFPN}. However, it is clear that a hybrid precoder or a phased array design is different from an all digital MIMO/massive MIMO implementations. Designing algorithms and bounds for hybrid precoding system while ignoring the implications of RF signal processing will lead to significant difference between achievable rates in theory and practice.

For example, to reduce number of RF chains, it is reasonable to have power amplifiers before RFPN (as shown in Fig. \ref{fig:1} and \cite{Elayach:hybrid}). However, this leads to a fundamental problem - signals with unequal amplitude and phase are combined in the RFPN, leading to signal dependent mismatch at the antenna ports, eventually resulting in significant power loss. Additionally, RFPN operating in millimeter wave regime will have non-negligible insertion loss \cite{Pozar:microwave}. Indeed, we must acknowledge that designing hybrid analog digital precoders requires a good understanding of the RF challenges as well as digital capabilities - an aspect that is often ignored in literature.

Recently \cite{Vijay:hybridmtt} propose algorithm, microwave designs and measurement results of some simple hybrid precoders used in cellular networks, and  show that hybrid precoders designed without considering RF signal processing limitations can lead to significant loss in effective radiated power. Subsequently, \cite{Adrian:hybrid} extends these insights to include RF system and  wireless channel models in order to quantify the impact of losses in hybrid precoders.

\subsection{Contributions and Outline}
In this paper, we start with the RF system models derived from hybrid precoder implementations \cite{Vijay:hybridmtt, Adrian:hybrid} and specify degradation from theoretical bounds for varying hybrid precoder architectures. In particular, some of the fundamental contributions of this work can be summarized as follows:
\begin{enumerate}
\item We show that the microwave transfer function of hybrid precoder is a coarse quantized representation of the ideal RF precoding matrix where the quantization resolution depends on the number of antennas and transceivers.
\item We derive a closed form expression to quantify loss in mutual information based on the microwave system model, and state the implications for realistic hybrid precoders.
\item We apply the above expression for various hybrid precoders such as a fully connected RFPN, sub-array RFPN and comment on the achievable rates and multi-user capacity.
\end{enumerate}
The rest of the paper is organized as follows: Sec. II starts with the RF system model \cite{Vijay:hybridmtt, Adrian:hybrid} and represents the microwave precoder as a quantized representation of ideal hybrid precoder. Sec. III derives analytical expressions on observed information loss due to the introduction of microwave/RF components in a hybrid precoders and analyses their impact for different hybrid precoders. Sec. IV shows simulation results followed by concluding remarks.

\emph{Notations}: Bold lower case and upper case letters denote vectors and matrices, while $\bzeros_{M}$ and $\bones_{M}$  denote $M$-element column vectors of zeros and ones respectively. Superscripts $(.)^{T}, \, (.)^{*}, \, (\hat{.}), \, (.)^{\dagger}$ and $\|.\|^{2}$ denote transpose, Hermitian transpose, estimate, pseudo-inverse, and Frobenius norm, respectively. Operation $\otimes$ denotes Kronecker products whereas $\vec$ operation transforms a matrix into a vector. An $n$-dimensional complex vector is denoted by $\mathbb{C}^{n\times 1}$, while $\mathbb{C}^{n\times m}$ denotes the generalization to an $(n \times m)$-dimensional complex matrix.

\bef
\begin{center}
\includegraphics[width=0.5\textwidth]{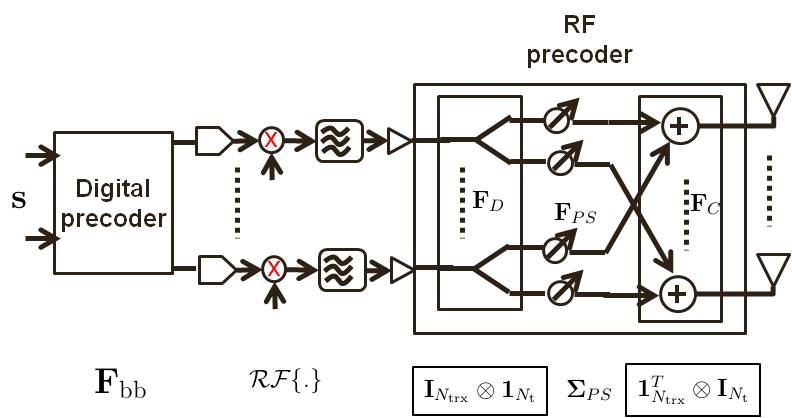}
\end{center}
\caption{\footnotesize{Base station employing hybrid precoder with $\Nx{trx}$ transceivers and $\Nx{t}$ antennas to communicate with users.}}
\label{fig:1}
\enf§

\section{RF System Model and Quantized Representation of RF Precoder}
\label{sec:RfSystemModel}
In this section, we will present the signal and mmWave channel model for the single user system under consideration.

\subsection{Hybrid Precoder Model}
Consider a single-user mmWave system consisting of a \ac{BS} with $\Nx{t}$ transmit antennas and $\Nx{trx}$ transceiver chains transmitting precoded data to a \ac{UE}  equipped with $\Nx{r}$ receive antennas. Specifically, the BS is assumed to have a hybrid architecture as in \cite{Elayach:hybrid, Vijay:hybridmtt}, where $\Nx{trx} < \Nx{t}$, and hence the BS employs a hybrid analog-digital precoding for transmission. Without loss of generality, it is assumed  that the user has a fully digital architecture, and is capable of performing optimal decoding of the received signals. The received signals are written as
\beq
\by = \sqrt{\rho} \bH^{T}\bx + \bw \quad \mathrm{where}\quad \bx = \bFx{RF} \bFx{BB} \bs,
\label{eq:1}
\enq
where $\by \in \mathbb{C}^{\Nx{r} \times 1}$  is the received signal vector, $\rho$ denotes the average received power, $\bH \in  \mathbb{C}^{\Nx{t}\times \Nx{r}} $ is the channel  matrix, $\bx \in \mathbb{C}^{\Nx{t}\times 1}$ is the transmitted signal vector, and $\bw \in \mathbb{C}^{\Nx{r}\times 1}$ denotes i.i.d additive noise vector drawn from  $\cC\cN (0, \sigma_{\mathrm{n}}^2)$. In a hybrid precoder \cite{Elayach:hybrid, Vijay:hybridmtt}, the data stream $\bs \in \mathbb{C}^{\Nx{s}\times 1}$, is first precoded by a digital precoding matrix $\bFx{BB} \in \mathbb{C}^{\Nx{trx} \times \Nx{r}}$, and subsequently by an  \ac{RFPN} denoted by the matrix $\bFx{RF} \in \mathbb{C}^{\Nx{t} \times \Nx{trx}}$ as shown in Fig. 1.  Our objective is to understand the impact of $\bFx{RF}$, whose elements are made of microwave components that typically include power dividers, phase shifters and power combiners \cite{Pozar:microwave}. 

Following \cite{Vijay:hybridmtt, Adrian:hybrid}, we represent the RFPN for the fully connected hybrid precoder, proposed in  \cite{Elayach:hybrid}, in the form of different microwave elements:
\bds
\item  Matrix $\bFx{D}$ of size $\Nx{t} . \Nx{trx} \times \Nx{trx}$ represents power dividers that divides $\Nx{trx}$ signals from the digital baseband unit into $\Nx{t}\Nx{trx}$ output signals, and is typically implemented as Wilkinson power dividers \cite{Pozar:microwave}.
\item The diagonal matrix $\bFx{PS}$ of size $\Nx{t} . \Nx{trx} \times  \Nx{t} . \Nx{trx}$ represents the RF phase shifters to achieve the desired analog precoding.
\item The power combiner matrix of size $\Nx{t} \times \Nx{t} . \Nx{trx}$ is denoted as $\bFx{C}$, and this matrix couples the phase shifted signals to the antenna array, and is typically implemented as hybrid directional couplers  \cite[Ch. 9]{Pozar:microwave}.
\eds
such that
\[
\bFx{RF} = \bFx{C} \bFx{PS} \bFx{D}.
\]
For simplicity, we assume that the power dividers and power combiners are balanced (i.e., of equal weights), and we also assume that the interconnections within the overall RFPN are kept fixed, and it is straightforward to extend these assumptions to arbitrary combination of microwave elements. Following \cite{Vijay:hybridmtt, Adrian:hybrid}, the S-parameter implementation of $\bFx{D}$ using Wilkinson power dividers can be written as
\[
\bFx{D} = \sqrt{\frac{1}{L_{\mathrm{s}}\Nx{t}}}\underbrace{ \left[\bea{cccc}
\bones_{\Nx{t}} &\bzeros_{\Nx{t}} &\cdots &\bzeros_{\Nx{t}} \\
\bzeros_{\Nx{t}} & \bones_{\Nx{t}} & \ddots  & \bzeros_{\Nx{t}} \\
\ddots & \ddots & \ddots  & \ddots \\
\bzeros_{\Nx{t}} & \bzeros_{\Nx{t}} & \ddots  & \bones_{\Nx{t}}
\ena\right]}_{ \Nx{t} . \Nx{trx} \times \Nx{trx}} = \left[ \bI_{\Nx{trx}} \otimes \bones_{\Nx{t}}\right]
\]
where $L_{\mathrm{s}}$ represents the substrate and implementation loss of power divider components. Likewise, $\bFx{PS}$ and $ \bFx{C}$ can be respectively represented as
\[
\bFx{PS} = \sqrt{\frac{1}{L_{\mathrm{ps}}}}\underbrace{ \left[\bea{cccc}
\sigma_{1,1} & 0 & \ddots & 0 \\
0 & \sigma_{1, 2} & 0 & \ddots\\
\ddots & \ddots & \ddots  & \ddots \\
0 & 0 & \ddots  & \sigma_{\Nx{t}, \Nx{trx}}
\ena\right]}_{ \Nx{t} . \Nx{trx} \times \Nx{t} . \Nx{trx}}
\triangleq  {\bSigma}_{\mathrm{PS}}
\]
and
\[
\bFx{C} = \sqrt{\frac{1}{L_{\mathrm{c}}\Nx{trx}}}\underbrace{ \left[\bea{cccc}
\bones_{\Nx{t}}^T & 0 & \ddots & 0 \\
0 & \bones_{\Nx{t}}^T & 0 & \ddots\\
\ddots & \ddots & \ddots  & \ddots \\
0 & 0 & \ddots  & \bones_{\Nx{t}}^T
\ena\right]}_{  \Nx{t} \times \Nx{t} \Nx{trx}}
 = \left[ \bones_{\Nx{trx}}^T \otimes  \bI_{\Nx{t}}\right].
\]
where $L_{\mathrm{ps}}$ and $L_{\mathrm{c}}$ denote the substrate loss in phase shifters and power combiners, respectively, and $\sigma_{i,i}$ denotes the phase of the  \ac{RFPN} matrix $\bFx{RF}$ \cite{Adrian:hybrid}. The aforementioned model of the fully connected RFPN can be further extended such that $\bF_{c}$ is made of a 4-port hybrid coupler \cite{Pozar:microwave}. One well known hybrid coupler based RFPN is the Butler matrix  \cite{Butler:butler_matrix}.
\subsubsection*{Subarray RFPN} When the RFPN has a subarray-based arrangement as in \cite{Vijay:hybridmtt, Adrian:hybrid}, then
\[
\bFx{RF} = \bFx{PS} \bFx{D}
\]
where the power divider matrix $\bFx{D}$ is of size $\Nx{t}  \times \Nx{trx}$ matrix given as
\[
\bFx{D} = \sqrt{\frac{1}{L_{\mathrm{s}}\frac{\Nx{t}}{\Nx{trx}}}}\underbrace{ \left[\bea{cccc}
\bones_{\frac{\Nx{t}}{\Nx{trx}}}  &\cdots &\bzeros_{\frac{\Nx{t}}{\Nx{trx}}} \\
\ddots &   \ddots  & \ddots \\
\bzeros_{\frac{\Nx{t}}{\Nx{trx}}} &  \ddots  & \bones_{\frac{\Nx{t}}{\Nx{trx}}}
\ena\right]}_{ \Nx{t}  \times \Nx{trx}},
\]
while $\bFx{PS}$ is an $\Nx{t}  \times \Nx{t}$ diagonal matrix. Note that there are no combiners in subarray arrangement.

\subsection{Saleh-Valenzuela Channel Model}
In this paper, we consider a discrete-time narrow-band Saleh-Valenzuela channel model as in \cite{Elayach:hybrid}, which is given by
\begin{eqnarray} \label{eq:ChannelModel}
\bH^{T} = \sqrt{\frac{\Nx{t}\Nx{r}}{L}} \sum_{l=1}^{L} \alpha_{l} \mathbf{a}_{\mathrm{r}}(\phix{l}{r}\thetax{l}{r}) \mathbf{a}_{\mathrm{t}}(\phix{l}{t}\thetax{l}{t})^{*},
\end{eqnarray}
where $L$ denotes the number of rays or paths from the \ac{BS} to the \ac{UE}, $\alpha_{l}$ is the complex gain associated with the $l$th ray or path, $\phix{l}{r}$ $(\thetax{l}{r})$ and $\phix{l}{t}$ $(\thetax{l}{t})$ are the azimuth (elevation) angles of arrival and departure for the $l$th path, respectively. The vectors $\mathbf{a}_{\mathrm{t}}(\cdot)$ and $\mathbf{a}_{\mathrm{r}}(\cdot)$ denote the  transmit and receive antenna steering responses, respectively, which depend on the antenna array used. We refer to \cite{Elayach:hybrid} for the array responses corresponding to a uniform planar antenna array and a uniform linear antenna array.

\subsection{Microwave approximation of ideal  $\bFrf$}
In the sequel, we show that the \ac{RFPN} realized by an arbitrary microwave implementation of $\bFx{C} \bFx{PS} \bFx{D}$ is a poor quantized representation of $\bFx{RF}$. To this end, we first assume that the microwave implementation of the \ac{RFPN} represented as $\bFx{C} \bFx{PS} \bFx{D}$ can be made sufficiently close to the ideal $\bFx{RF}$ in terms of chordal distance between them on the Grassmannian manifold. Based on this assumption, we approximate this chordal distance as the Euclidean distance between the \emph{ideal} $\bFx{RF}$ in \cite{Elayach:hybrid} and the microwave realization $\bFx{C} \bFx{PS} \bFx{D}$ \cite{Vijay:hybridmtt} as  $\|\bFx{RF} - \bFx{C} \bFx{PS} \bFx{D}\|_{F}$ with
\beq
\bFx{C} \bFx{PS} \bFx{D} =\left[ \bones_{\Nx{trx}}^T \otimes  \bI_{\Nx{t}}\right]
 \,  {\bSigma}_{\mathrm{PS}} \,
\left[ \bI_{\Nx{trx}} \otimes \bones_{\Nx{t}}\right]
\label{eq:2}
\enq

Vectorizing the above matrix product using the identity $\vec(\bFx{C} \bFx{PS} \bFx{D}) = (\bFx{D}^T \otimes \bFx{C}) \vec ( \bFx{PS} )$ \cite{Horn:matrix_anal} yields
\[
\bea{@{}c@{}c@{}c@{}}
\vec(\bFx{C} \bFx{PS} \bFx{D}) & =&
\left[ \left[ \bI_{\Nx{trx}} \otimes \bones_{\Nx{t}}\right]^{T} \otimes \left[ \bones_{\Nx{trx}}^T \otimes  \bI_{\Nx{t}}\right] \right] \vec( {\bSigma}_{\mathrm{PS}}), \\
 &=& \und{\bP}  \, \und{\bsigma},
\ena
\]
where $\und{\bP}$ is of size $N_{\mathrm{trx}}N_{\mathrm{t}} \times N_{\mathrm{trx}}^2N_{\mathrm{t}}^2$, and $\und{\bsigma}$ is of size $N_{\mathrm{trx}}^2 N_{\mathrm{t}}^2 \times 1$, which is written as
\[
\und{\bsigma} =
\left[\bea{ccccc} \sigma_{1,1}, \bzeros_{\Nx{t}\Nx{trx}}, \hdots, \sigma_{\Nx{t}, \Nx{trx}}, \bzeros_{\Nx{t}\Nx{trx}}
\ena
\right].
\]
A few remarks are in order regarding the microwave approximation of \ac{RFPN}:
\bds
\item[{[R1]}] Note that in the above expression, $\vec(\bFx{RF})$ has $\Nx{t}\Nx{trx}$ non-zero entries along the unit circle. Similarly, ${\und{\bsigma}}$ has $\Nx{t}\Nx{trx}$ non-zero entries, and a one-to-one mapping between $\vec(\bFx{RF})$ and ${\und{\bsigma}}$ is feasible, as long as $\und{\bP}$ is a tall matrix with full column rank.
\item[{[R2]}] An inspection of $\|\vec(\bFx{RF}) -  \und{\bP}  \, \und{\bsigma} \|_{F}$ reveals that  $\und{\bP}$ is a fat matrix as  $N_{\mathrm{trx}}N_{\mathrm{t}} \ll N_{\mathrm{trx}}^2N_{\mathrm{t}}^2$. Additionally,  $\und{\bP}$ is a \emph{sparse} matrix with each row containing only $N_{\mathrm{trx}}N_{\mathrm{t}}$ non-zero entries among $N_{\mathrm{trx}}^2N_{\mathrm{t}}^2$ entries. Thus  $\und{\bP}  \, \und{\bsigma} = \vec(\bFx{RF}) $ represents an under-determined system of linear equations in $\und{\bsigma}$, implying that irrespective of the resolution of phase shifters, there will be deviations induced between $\bFx{RF}$, and the microwave realization of the \ac{RFPN}, i.e., $\bFx{C} \bFx{PS} \bFx{D}$.
\eds
We refer to the deviation $\|\bFx{RF} - \bFx{C} \bFx{PS} \bFx{D}\|_{F}$ as the quantization error. In other words, [R2] specifies that in a fully connected \ac{RFPN}, the quantization error will always be non-negligible. The quantization matrix $\ubP$ operates on the sparse vector $\und{\bsigma}$ whose non-zero entries are uniformly sampled as shown in (\ref{eq:2}). Our objective is to represent $\bFx{RF}$ in terms of the quantized representation $\und{\bP}  \, \und{\bsigma}$ while minimizing the quantization error. Note that in a given fully connected hybrid precoder, since $\bFx{D}$ and $\bFx{C}$ are fixed, our design choices are limited to sparse $\und{\bsigma}$.

From the structure of $\ubP$, for a given $\Nx{t}. \Nx{trx}$, it is clear that the row weight and column weight of $\ubP$ remains constant. Thus the quantization levels are same. For example, consider the $\ubP$ for different values of the tuple $(\Nx{t}, \Nx{trx})$, i.e., $\{(\Nx{t}, \Nx{trx}) := (64,2), (32,4), (16,8), \ldots \}$. Here the row and column weights for all $\ubP$'s will be the same. However, the distribution of the non-zero entries in $\ubP$ that quantize $\bFx{RF}$ will vary significantly. This has interesting implications, which are as follows.
\bds
\item[{[R3]}] By increasing $\Nx{t}$ and reducing $\Nx{trx}$, the non-zero entries in each row of $\ubP$ are more uniformly distributed. This intuitively implies that as $\Nx{t}$ increases, $\ubP$ uniformly quantizes the uniformly spaced entries in $\und{\bsigma}$. Conversely, by reducing $\Nx{t}$ and increasing $\Nx{trx}$, $\ubP$ non-uniformly quantizes $\und{\bsigma}$.
\item[{[R4]}] Since the non-zero elements in $\und{\bsigma}$ are uniformly spaced, uniform quantization due to $\ubP$ by increasing $\Nx{t}$ reduces the overall quantization error. Similarly, increasing $\Nx{trx}$ leading to non-uniform quantization with $\ubP$ will result in increasing quantization error.
\eds

\section{Rate Degradation in Realistic Hybrid RF Precoding Systems}
\label{sec:RateDegradation}

In the previous section, we observe that the constraints placed by the microwave representation of \ac{RFPN} leads to a quantized representation of $\bFx{RF}$, and that increasing $\Nx{trx}$ will lead to increased quantization error. In this section, we will analyze the impact of the microwave RFPN on the achievable rate. 

Our objective is to quantify the loss in mutual information when the quantized microwave representation is used at the transmitter for realizing a hybrid RF precoding. Note that algorithms to design \emph{ideal} $\bFx{RF}$ is not the focus of this work -- for more information, the readers are encouraged to refer to \cite{Elayach:hybrid, Vijay:hybridmtt, Lau:hybrid, Molisch:RF_phase_shifter}. Following the methodology in \cite{Elayach:hybrid}, we  design $\bFx{RF}$, and subsequently design $\bFbb$ as a function of $\bFrf$ and the unconstrained optimal precoder of the MIMO channel under consideration:
\[
\bFrf \bFbb = \bFopt \quad \textrm{or} \quad \bFbb = \bFrf^{\dagger} \bFopt.
\]

For the rest of the paper, we assume that $\bFx{BB}$ is ideal, and we are concerned with the impact of practical hybrid precoders on the mutual information between $\bs$ and $\by$: \newline \newline
$\cI(\bH, \bFx{D}, \bFx{PS}, \bFx{C})$
\beqa\label{eq:MutualInfn1}
\quad &=& \log_2 	
\left[
\left| \bI + 
 \gamma_s \,
\bH \bFx{C} \bFx{PS} \bFx{D} \bFx{BB}(\bH \bFx{C} \bFx{PS} \bFx{D} \bFx{BB})^{*} \right|
\right]\nonumber\\
 &=& \log_2 	
\left[  \left| \bI +  \gamma_s \, .   \right. \right.\nonumber\\
& & \quad  \left.\left. {\bSigma}_{\mathrm{H}}^2 \bV^{*} \bFx{C} \bFx{PS} \bFx{D} \bFx{BB}( \bFx{C} \bFx{PS} \bFx{D} \bFx{BB})^{*} \bV \right|
\right]
\enqa
where $\gamma_s = \frac{\rho}{\Nx{s}\sigmax{n}^2}$. %
Here the SVD of $\bH$ is given as $\bH = \bV  {\bSigma}_{\mathrm{H}} \bU^{*}$. Also, the receiver operation has been abstracted by focusing on the hybrid precoder, under the assumption that the receiver is capable of performing optimal decoding based on the received signal $\by$. Now, exploiting sparsity in $\bH$, ${\bSigma}_{\mathrm{H}}$ is partitioned as
\[
 {\bSigma}_{\mathrm{H}}= \left[\bea{cc}
\bSigma_{1} & 0 \\
0 & \bSigma_{2}
\ena\right],  \bV= \left[\bea{cc}
\bV_{1}, \bV_{2} \ena\right],
\]
where $\bV_{1}$ and $\bSigma_{1}$ are $\Nx{t} \times \Nx{s}$ and $\Nx{s} \times \Nx{s}$ matrices, respectively. Since $\bV_{1}$ represents the basis vectors that span the signal sub-space of the channel matrix $\bH$ (and $\bV_{2}$ spans the null space of $\bH$), the optimal unconstrained precoder is simply given by $\bV_{1}$. However, this is not generally applicable for a hybrid precoder as correctly pointed out by \cite{Elayach:hybrid}. Some remarks on approximating $\bV_{1}$ using $\bFx{C} \bFx{PS} \bFx{D} \bFx{BB}$ are:
\bds
\item[{[R5]}] One of the important conclusions of the analysis of large scale uniform linear arrays in joint spatial division multiplexing (JSDM) \cite{Adhikary:jsdm} is that for large values of $\Nx{t}$, the basis vectors in the channel correlation expression can be approximated by columns of discrete Fourier transformation (DFT) matrix.
\item[{[R6]}] This suggests that the prebeamforming matrix in \cite{Adhikary:jsdm} or RFPN in our work can be obtained by choosing $\Nx{trx}$ of the $\Nx{t}$ columns from the DFT matrix. As shown in \cite{Butler:butler_matrix}, RF beamforming networks can be efficiently designed using branch hybrid couplers, such that either $\bFc$ or the overall RFPN, $\bFx{C} \bFx{PS} \bFx{D}$, can represent $\Nx{trx}$ columns of a DFT matrix.
\item[{[R7]}] Indeed, higher order RFPN's can be designed \cite{Shelton:fft_butler, Vijay:hybridmtt} such that $\Nx{trx}$ columns of the $\Nx{t}$-order DFT matrix is used as the prebeamforming matrix while ensuring that for large values of $\Nx{t}$, the overall product $\bFx{C} \bFx{PS} \bFx{D} \bFx{BB}$ can be made sufficiently close to $\bV_{1}$.
\eds

Based on the comments [R5-R7], we approximate the unitary property associated with $\bV_{1}$ as $\bV_{1}^*\bFx{C} \bFx{PS} \bFx{D} \bFx{BB} \approx \mathbf{I}_{\mathrm{\Nx{s}}}$. Consequently, we have that $\bV_{2}^*\bFx{C} \bFx{PS} \bFx{D} \bFx{BB} \approx \mathbf{0}$. Then the mutual information in \eqref{eq:MutualInfn1} can be simplified as \newline \newline
$\cI(\bH, \bFx{D}, \bFx{PS}, \bFx{C})$
\beqa\label{eq:MutualInfn2}
\quad  &=& \log_2 		
\left[
\left| \bI + \gamma_s \, .   \right. \right.\nonumber\\
& & \quad  \left.\left.
 {\bSigma}_{\mathrm{H}}^2 \bV^{*} \bFx{C} \bFx{PS} \bFx{D} \bFx{BB}( \bFx{C} \bFx{PS} \bFx{D} \bFx{BB})^{*} \bV \right|
\right]\nonumber\\
&=& \log_2 		
\left[
\left| \bI + \gamma_s \left[\bea{cc}
\bSigma_{1}^2 & 0 \\
0 & \bSigma_{2}^2
\ena\right] 
\, .   \right. \right.\nonumber\\
& & \quad  \left.\left.\bV^{*} \bFx{C} \bFx{PS} \bFx{D} \bFx{BB}( \bFx{C} \bFx{PS} \bFx{D} \bFx{BB})^{*} \bV \right.\Bigg|
\right.\Bigg] \nonumber\\
&\approx& \log_2 		
\left[
\left| \bI + \gamma_s  \,   \right. \right.\nonumber\\
& & \quad  \left.\left.
{\bSigma}_{1}^2 \bV_{1}^{*} \bFx{C} \bFx{PS} \bFx{D} \bFx{BB}( \bFx{C} \bFx{PS} \bFx{D} \bFx{BB})^{*} \bV_{1} \right|
\right] \nonumber\\
&=& \log_2 		
\left[\left|\bI +\gamma_s  \,   {\bSigma}_{1}^2\right|\right] + \log_2
\left[ \left| \bI - \left( \bI + \gamma_s  \, .  {\bSigma}_{1}^2 \right)^{-1} \gamma_s  \, .\right.\right. \nonumber\\
&& \left.\left.  {\bSigma}_{1}^2  \left[ \bI -  \bV_{1}^{*} \bFx{C} \bFx{PS} \bFx{D} \bFx{BB}( \bFx{C} \bFx{PS} \bFx{D} \bFx{BB})^{*} \bV_{1} \right]  \Big|
\Big]\right.\right., \nonumber \\
&&
\enqa
where the second term in \eqref{eq:MutualInfn2} represents the loss in mutual information due to the quantization error induced by the microwave implementation of the \ac{RFPN}.

We now represent $\bFx{BB}$ in terms of the $\bV_{1}$ and $\bFx{RF}$ as
\[
\bFx{BB} = (\bFx{RF}^{*}\bFx{RF})^{-1}\bFx{RF}^{*} \bV_{1} \quad \textrm{where} \quad \bFx{RF} = \bFx{C} \bFx{PS} \bFx{D}
\]
{leading to}
\[
\bFx{RF}\bFx{BB} = \bFx{RF} \left[\bFx{RF}^{*} \bFx{RF} \right]^{-1} \bFx{RF}^{*} \bV_{1} = \bP_{\mathrm{RF}} \bV_{1}
\]
\textrm
{where }
\[
\bP_{\textrm{RF}} = \bFx{C} \bFx{PS} \bFx{D} \left[(\bFx{C} \bFx{PS} \bFx{D})^{*} (\bFx{C} \bFx{PS} \bFx{D}) \right]^{-1} (\bFx{C} \bFx{PS} \bFx{D})^{*}
\]
 is an low rank projection matrix of size $\Nx{t} \times \Nx{t}$. After some algebraic manipulation, the mutual information in \eqref{eq:MutualInfn1} can be rewritten as
\newline \newline
$\cI(\bH, \bFx{D}, \bFx{PS}, \bFx{C})$
\beqa
\quad & =&  \log_2 		
\left[\left|\bI + \gamma_s  {\bSigma}_{1}^2\right|\right] + \log_2
\left[ \left| \bI - \left( \bI + \gamma_s  {\bSigma}_{1}^2 \right)^{-1}  \right.\right. \nonumber \\
& & \left. \left. \gamma_s   {\bSigma}_{1}^2  \left[ \bI -  \bV_{1}^{*}  \bP_{\textrm{RF}} \bV_{1} \right]  \Big|
\Big]\right.\right.
\label{eq:MutualInf3}
\enqa
\bds
\item  In the above expression, projection matrix $\bP_{\textrm{RF}}$  results in dimensionality reduction leading to a loss in mutual information. This means that as we progress from \emph{ideal} hybrid precoding techniques to their realistic implementations, the loss in mutual information is depends on the microwave realization of RFPN. 
\eds
The quantization error in (3) manifests as loss in mutual information and primarily limits the performance of realistic fully connected hybrid precoders. In the ensuing section, we present simulation results that confirm the accuracy of our analytical findings.

\section{Simulation Results}
\label{sec:Simulations}

In this section, the analytical results presented in Section \ref{sec:RateDegradation} are verified by comparing them against the results obtained from Monte Carlo simulations. We consider a system that consists of a single cell with a BS having $\Nx{t}=256$ antennas, and a \ac{UE} with $\Nx{r}=16$ antennas. For simplicity, we keep $\Nx{r} = \Nx{s}$for simulations. The centre frequency considered is $30$ GHz. The BS is assumed to have a hybrid precoder with $\Nx{trx} = 8$ RF chains, while optimal decoding is assumed at the user. Further, we assume that $6$ bits are used to quantize the steering angles at the transmitter when the ideal hybrid precoder from \cite{Elayach:hybrid} is considered. For the microwave implementation of the sub-array based precoder, and the fully connected precoder, $\bFx{BB}$ is designed as a zero-forcing precoder based on the effective channel formed by cascading the wireless channel $\bH$ with the RFPN.

The channel between the BS and the UE is modelled as a $10$-ray channel with random angles of arrival and departure \cite{Elayach:hybrid}. Specifically, the gain of each of the ten paths between the BS and the UE is drawn from a normal Gaussian distribution, while the angles of arrival and departure are drawn from a Laplacian distribution as in \cite{Elayach:hybrid}. The angular spread at the transmitter and the receiver is kept small at $10^{\circ}$, given that the channel is a mmWave channel. We assume that the BS uses a uniform planar antenna array, while a uniform linear antenna array is used at the UE. Also, it is assumed that the BS has perfect knowledge of the channel $\bH$. A total power constraint is applied for all the precoding methods considered, and the SNR is defined as $\frac{\rho}{\sigmax{n}^2}$. For downlink transmission, equal power allocation across the BS antennas is considered.

\begin{figure}[!t]
\begin{center}
\includegraphics[width = 3.75in, keepaspectratio=true]{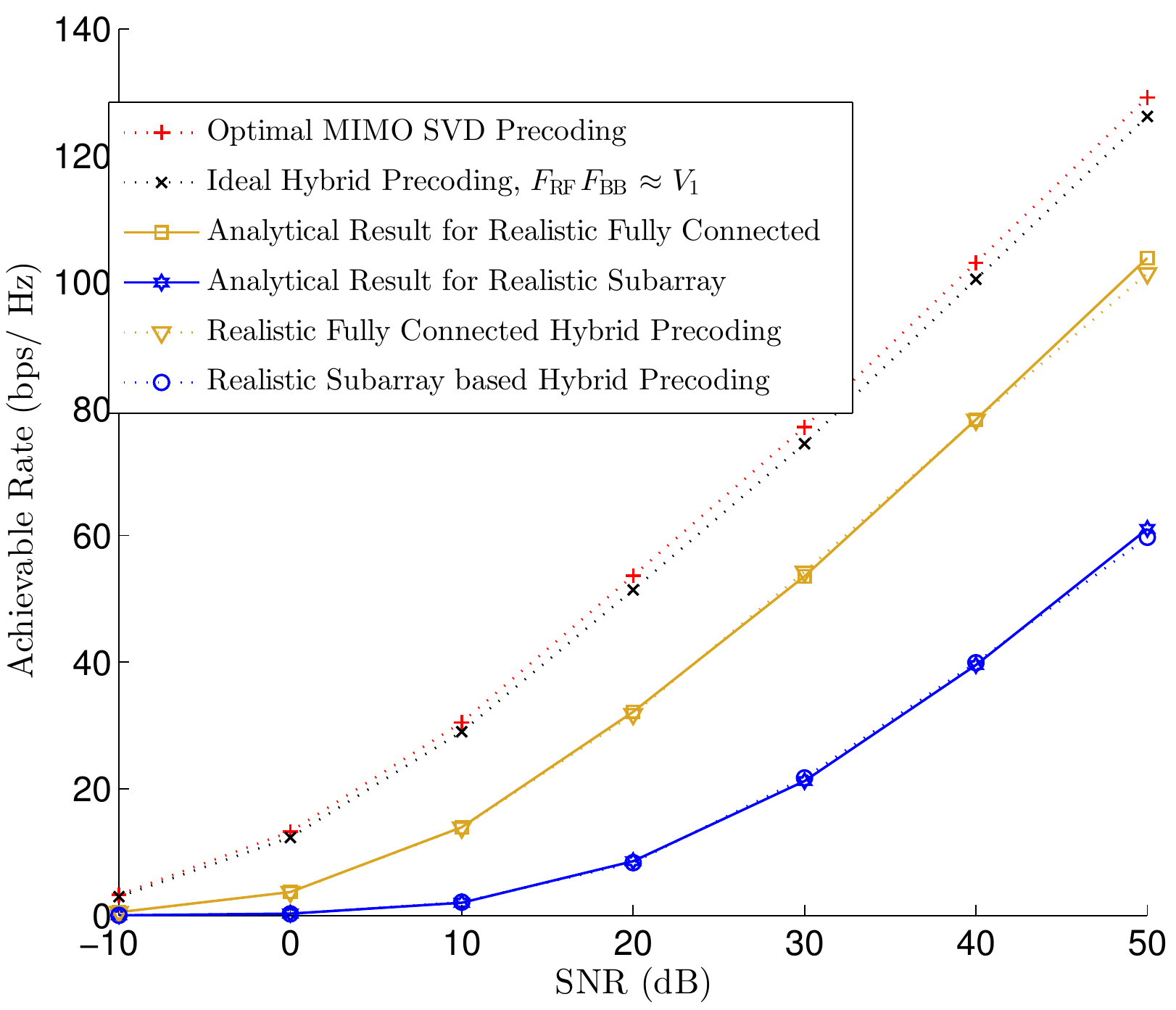}
\caption{Comparison of achievable rates for the different hybrid precoding methods for $\Nx{t}=256$, $\Nx{r}=16$, and $\Nx{trx} = 8$.  }
\label{fig:NTx256NRf8}
\end{center}
\end{figure}

\begin{figure}[!t]
\begin{center}
\includegraphics[width = 3.75in, keepaspectratio=true]{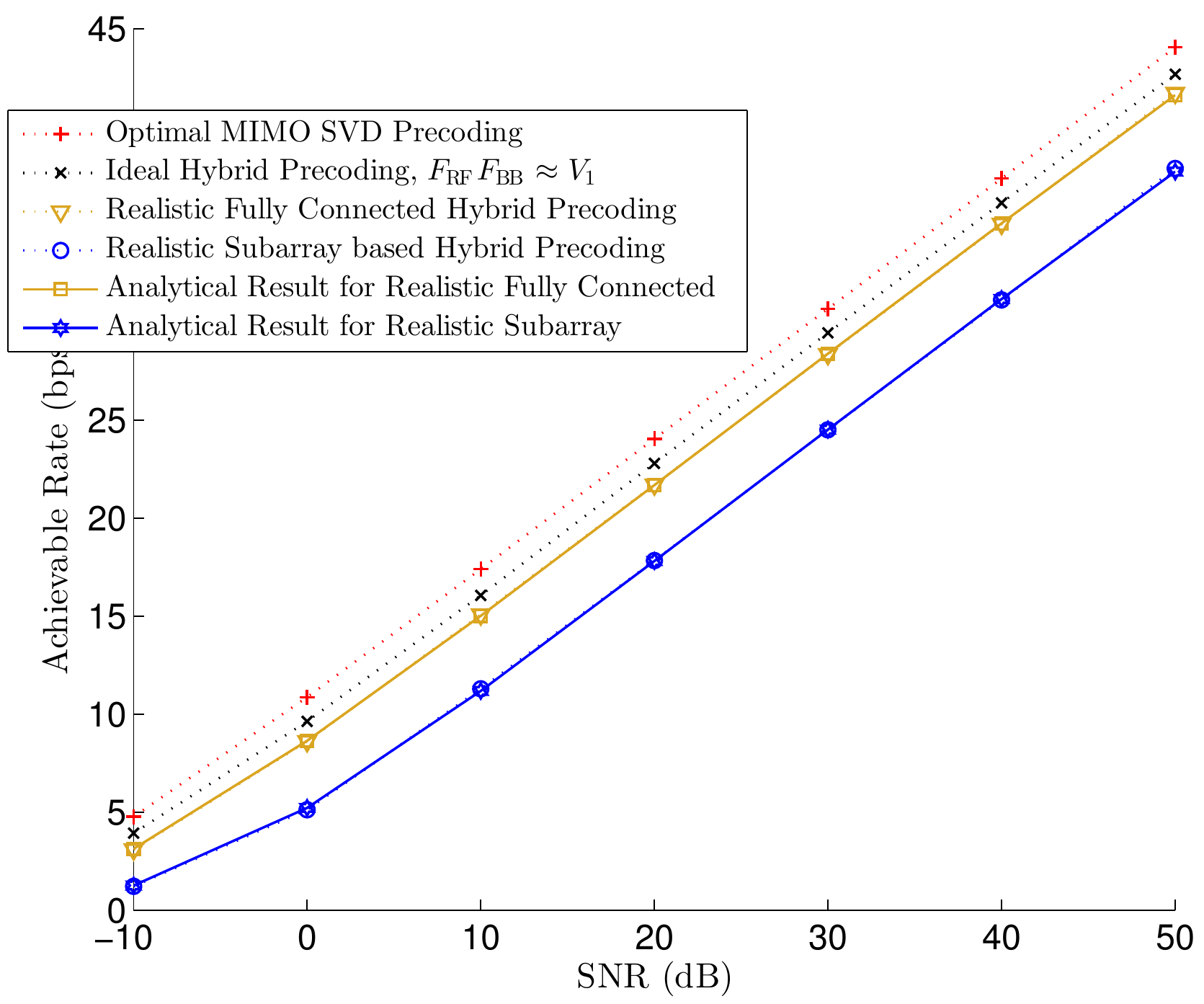}
\caption{Comparison of achievable rates for the different hybrid precoding methods for $\Nx{t}=256$, $\Nx{r}=16$, and $\Nx{trx} = 2$.  }
\label{fig:NTx256NRf2}
\end{center}
\end{figure}

In Figs. \ref{fig:NTx256NRf8}, and \ref{fig:NTx256NRf2} (where $\Nx{trx}=2$ is used), we observe that the theoretical mutual information derived for the realistic fully connected and sub-array based hybrid precoding methods in \eqref{eq:MutualInfn2} closely match the simulation results. This verifies the accuracy of the analytical results derived. Furthermore, similar to the result in \cite{Elayach:hybrid}, we observe that the achievable rates for the theoretical hybrid precoding method is close to that of the optimal digital SVD based precoding. However, for the realistic implementation of the fully connected hybrid precoding method, a significant degradation of performance is observed for the realistic fully connected hybrid precoding method. Furthermore, the performance of the realistic sub-array-based precoding is seen to be much lower than that of the realistic fully connected  precoding method.

As a confirmation of [R3], we observe that the degradation in performance due to the microwave implementation of the phase shifting network increases with increase in $\Nx{trx}$ for a given $\Nx{t}$. Specifically, the gap in performance between the realistic methods and the ideal hybrid precoding method decreases with increase in $\Nx{trx}$ for a given $\Nx{t}$.  This observation is in line with the discussion in Section \ref{sec:RateDegradation}.

\begin{figure}[!t]
\begin{center}
\includegraphics[width = 3.75in, keepaspectratio=true]{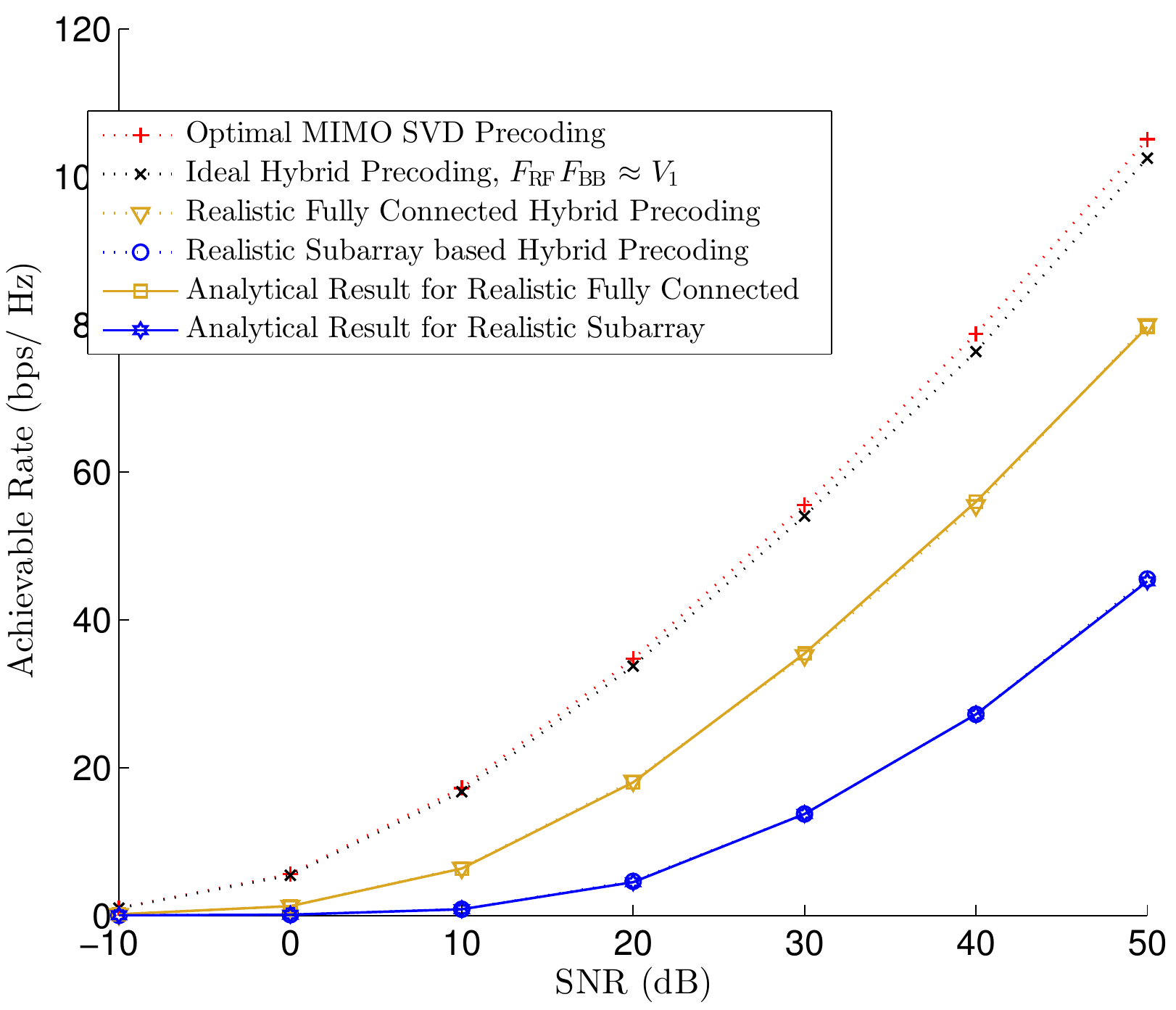}
\caption{Comparison of achievable rates for the different hybrid precoding methods for $\Nx{t}=64$, $\Nx{r}=16$, and $\Nx{trx} = 2$.  }
\label{fig:NTx64NRf8}
\end{center}
\end{figure}

We consider $\Nx{t} = 64$ and $\Nx{trx} = 8$ in Fig. \ref{fig:NTx64NRf8}, and upon comparing with the result in Fig. \ref{fig:NTx256NRf8}, for a given  $\Nx{trx}$, we observe that the relative degradation in the performance of the realistic implementation decreases with increase in $\Nx{t}$. For instance, at medium SNR regime i.e. SNR$=10$ dB, the performance degradation of the realistic fully connected array for $\Nx{t}=256$ is about $50$$\%$ of that of the ideal hybrid precoder, while for $\Nx{t}=64$, the degradation of the realistic fully connected array is about $35$$\%$ of that of the ideal hybrid precoder. At the high SNR regime, SNR$=50$ dB, the performance of the realistic fully connected array for $\Nx{t}=256$ is about $19$$\%$ of that of the ideal hybrid precoder, while for $\Nx{t}=64$, the performance of the realistic fully connected array is about $24$$\%$ of that of the ideal hybrid precoder.

\section{Conclusion}
We show that practical hybrid precoding methods exhibit a significant performance degradation when realistic microwave implementations are considered. Furthermore, we observe that the performance  degrades with increase in the number of transceiver chains, due to increase in the quantization error involved in representing $\bFx{RF}$ using realistic microwave elements. On one hand, they indicate that we must redesign RFPNs by accounting for RF effects and limitations. On the other hand, they provide pointers that for large scale arrays, RFPNs designed to approximate DFT matrices can lead to reasonable performance. 

\bibliographystyle{ieeetran}
\bibliography{mybibliography}
\end{document}